# Uncovering the epistemological and ontological assumptions of software designers

***David KING***
*Research Student*
University of York
Department of Computer Science,
+44 1904 432749
**dk@cs.york.ac.uk**

***Chris KIMBLE***
*Lecturer*
University of York
Department of Computer Science,
+44 1904 433380
**kimble@cs.york.ac.uk**

**Résumé**

Les positions ontologiques et épistémologiques adoptées par des méthodes de conception de systèmes d'information sont incommensurables une fois prises à leurs extrémités. Les chercheurs dans des systèmes d'information tendent donc à se concentrer sur les *similitudes* entre différentes positions, habituellement à la recherche d'un simple, position d'unification. Cependant, en se concentrant sur les similitudes, la clarté de l'argument fournie par n'importe quelle une position philosophique unique est nécessairement diminuée. En conséquence, les chercheurs traitent souvent les bases philosophiques des méthodes de conception en tant qu'étant d'intérêt seulement minoritaire. En cet article, nous avons délibérément choisi de concentrer sur les *différences* entre de diverses positions philosophiques. Aller de cette position, nous croyons que nous pouvons obtenir une connaissance plus claire du comportement empirique du logiciel comme vu des positions philosophiques particulières. Puisque l'évidence empirique ne favorise aucune position unique, nous concluons cela les approches ad-hoc fournissent plus forte et plus théoriquement enraciné approche à la conception de logiciel.

**Mots clefs:**
Méthodes de Conception de Logiciel, Théorie Philosophique, Équivalence, Epistémologie, Ontologie.

**Abstract**

The ontological and epistemological positions adopted by information systems design methods are incommensurable when pushed to their extremes. Information systems research has therefore tended to focus on the *similarities* between different positions, usually in search of a single, unifying position. However, by focusing on the similarities, the clarity of argument provided by any one philosophical position is necessarily diminished. Consequently, researchers often treat the philosophical foundations of design methods as being of only minor importance. In this paper, we have deliberately chosen to focus on the *differences* between various philosophical positions. From this focus, we believe we can offer a clearer understanding of the empirical behaviour of software as viewed from particular philosophical positions. Since the empirical evidence does not favour any single position, we conclude by arguing for the validity of ad hoc approaches to software design which we believe provides a stronger and more theoretically grounded approach to software design.

## Preface

Philosophy should be no stranger to information systems. After all, the body of theory underlying programs originated from work undertaken in the philosophy of mathematics in the early 20$^{th}$ Century. Nonetheless, more recent work has often lost sight of these foundations leading to a blurring of the distinction between the term's 'software' and 'programs'. In this paper, we characterise *programs* as formal descriptions of the actions of a computer. The focus of programs is thus clearly on the actions of the computer — any description with a wider scope falls into the category of *software*.

The distinction between software and program is important, because it makes clear that software design and program design are distinct activities. Program design focuses on producing detailed descriptions of the activities within a computer; software design focuses on wider descriptions aimed at characterising problems and solutions both inside and outside the machine. Consequently, while software design includes program design, software design is *not* reducible to program design. The properties that are expected of programs in program design should *not* automatically be expected of software in software design.

Notwithstanding this, researchers in software design often use programming theory as the inspiration for their work. Programming theory however is bound to a certain philosophical assumptions. Although *software* design methods rarely share *all* the philosophical assumptions of *program* design methods, few design methods appear to see this as a problem. However, a problem does exist when designers unconsciously adopt philosophical positions unsupported by the design methods they are using. The behaviour expected of the world by the design method may not match the actual behaviour — if designers are unaware of this conflict, they may continue to use unsuitable design methods.

In this paper we argue that by un-picking the philosophical arguments, a broader framework can be established. This broader framework[1] moves away from a single, unified, view of Information Systems design. Instead, we argue for software design being underpinned by a series of theories that we describe as 'ad hoc', by which we mean, "*devoted, appointed to or for some particular purpose*" (Simpson J. and Weiner E. 1989). Using these ad hoc theories, allows us to understand more clearly the empirical behaviour of software development under these different philosophical assumptions. Thus, studying different ad hoc theories should lead to a better empirical understanding of the process of software design.

## 1. Properties of Descriptions

### 1.1 What is a Description?

In this paper, we concentrate on philosophical notions of epistemology and ontology. Since the philosophical notions often differ (Zúñiga G. 2001), we first need to define our terms[2].

To define what we mean by a 'description' we must start with a recognition that an *unperceived* universe exists: a universe beyond the reach of any individual designer. We will call this unperceived universe *reality*, defined as "...*that which underlies and is the truth of appearances or phenomena*" (Simpson J. and Weiner E. 1989). In common with other philosophical discussions, the emphasis in this discussion is on what really exists, rather than existence as perceived by any individual designer (Audi R. 1995). Notwithstanding this, the perceptions of an individual designer do matter and can be important to the success of the software design process (Sommerville I. and Sawyer P. 1997). For example, different users have different views of the system being designed; it is only by capturing these different viewpoints that a designer can fully understand the needs of the different users (Graham T. 1996). Thus in software design, in addition to unperceived reality, we need a notion of the designer's 'perceived reality'; we will call this 'perceived reality' the *representation*.

In essence, the representation can be thought of as the mental model that an individual designer uses to characterise perceived reality. The representation can therefore be defined as "[t]*he operation of the mind in forming a clear image or concept*" (Simpson J. and Weiner E. 1989). However, these mental models are internal to the designer and are not directly transmissible. Instead, some intermediate form is needed; a form we will call the *description*. We will simply define a description as "[a] *statement which describes, sets forth, or portrays*'' (Simpson J. and Weiner E. 1989).

Having offered a working definition for the terms *description*, *representation* and *reality*, we will now turn to the terms that link these concepts: *information* and *knowledge*. We will start with the relationship between reality and representation.

Essentially, this relationship characterises the representation as the result of the reception of stimuli from reality. Following Charles Meadow and Weijing Yuan (1997), we will refer to this as *information*. Philosophically, we view the properties of the information relationship as the concern of *ontology*. Ontological arguments are con-cerned with the *nature* of reality; without any concern for how that nature might be 'known'.

---

[1] This framework is described in more detail in the accompanying paper, King D. and Kimble C. (2004)

[2] We have chosen to use standard dictionary definitions where possible. Although we recognise that these definitions may imply a particular philosophical position, it is not our intention to advocate any single position here.

By contrast, the formation of a description from a representation is a conscious act; a designer knowingly chooses the form of the description. We will therefore refer to this relationship as *knowledge*. Philosophically, we view the properties of the knowledge relationship as the concern of *Epistemology*. Epistemological arguments are those that focus on the study and characterisation of knowledge.

## 1.2 The Use of Philosophy

All five terms defined in §1.1 are briefly summarised pictorially in Figure 1.

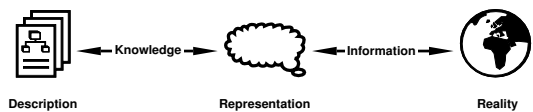

**Figure 1 –Summary of Definitions and Terms**

Although our choice of working definition for the term 'knowledge' may seem controversial, it is not our intention to enter into arguments about the various distinctions that can be made between the terms 'knowledge' and 'information' here. Instead of looking for a single universal definition of 'knowledge' and 'information' here, we are only concerned with the ad hoc notions of 'knowledge' and 'information' that software designers use in their work.

The aim of software design is to create descriptions that communicate characterisations of both a problem and its solution. Consequently, the properties of the description used by the software designer critically affect the communicational ability of the design. This means the success of the software rests on the ability of the designer to create rich, high-quality descriptions.

Most discussions of software design focus on the low-level properties of a particular style of description or method of software design. However, software descriptions also have broad, high-level properties that are often ignored. It is these high-level properties that most affect a designer's ability to create rich, high-quality descriptions and it is these properties that are discussed in this paper. To discuss these properties of software descriptions more fully we must turn to philosophy.

## 1.3 Philosophy and design

In this section, we will outline four broad approaches to software design and discuss their philosophical underpinnings. These approaches and their associated philosophical underpinning are summarised in Table 1 overleaf. We will begin our discussion with the oldest and arguably the most influential of all methods: formal methods.

Historically, formal methods arose from the debates of the rationalist school of philosophy. Rationalist arguments deal principally with epistemology[3], and broadly claim, "*Reason is the source of all knowledge*". Hence, "*[i]f reason is the source of all knowledge, then everything that can be known including the natural world must be intelligible and rationally explicable*" (Jack A. 1993). As Jack notes, most rationalist positions argue that not only can all knowledge be uncovered by rational thought, but so too can the properties of the natural world. This latter point is important, because rationalist epistemological arguments are often tied to *realist* ontological[4] arguments. Realism "*emphasises that truth is possible: beliefs are testable against 'reality', and that reality is 'knowable'* " (Duro P. *et. al.* 1993).

Design methods that argue from realist ontologies and rationalist epistemologies form the **formal** strand of research in software design. Here there is a seamless equivalence between the software description, the representation in the designers mind and the underlying aspects of reality that are being modelled.

In practice however, most design methods choose to abandon either the assumption of rationality, or the assumption of realism. Thus, although common in *program* design methods, few *software* design methods adhere fully to the principals of the formal design methods. Nevertheless, these principals have been, and remain, hugely influential.

As indicated above, the earliest software design methods were derived directly from program design methods. However, as many researchers have noted, the translation between description and reality is not a trivial problem. Consequently, many software designs methods have chosen to recognise, sometimes only implicitly, a distinction between logical and physical designs.

Software designs that were *logically* correct do not always reflect the properties the same software design has in reality. Consequently, software design methods in the **semi-formal** strand try to *manage* the divide between logical and physical design rather than ignoring it.

Recognising a split between logical and physical designs means basing software design methods on rationalist epistemological arguments and *anti-realist* ontological arguments. Anti-realist arguments claim that the perception of reality is so bound to the mind observing reality that (Kant I. 1996):

*Even if we could impart the highest degree of clearness to our intuition, we should not come one step nearer to the nature of objects by themselves.*

---

[3] Epistemology is concerned with theories of knowledge, asking: "What can we know and how do we know it"?

[4] Ontology is concerned with theories of existence, asking: "What is the essence and nature of the world"?

Anti-realist arguments therefore reject the assumption that logical properties of a description are *guaranteed* to hold in reality.

However, most software design methods remain attracted to the notion of a seamless equivalence between the software description and the underlying reality. So rather than abandoning the realist ontological arguments, modern design methods more often abandon the rationalist epistemological arguments.

Design methods in the **object-oriented** strand argue for a reversal of the assumptions made by the semi-formal strand. Rather than working from the description *to* reality, work *from* reality to the description. Instead of using a rationalist epistemology, these design methods use an empiricist epistemology. Empiricism argues for knowledge as the result of observation and experience. Rather than deriving reality from a perfect, rational, description, empiricist design methods seek to deduce features of the description from reality.

Nonetheless, design methods in the object-oriented strand argue that once the description is formed from observation of reality, the description and reality will remain synchronised. Thus, the description can be manipulated to reveal properties of reality. Design methods in the object-oriented strand thus preserve realist ontologies.

Finally, a few design methods abandon both rationalist epistemologies and realist ontologies. For example, in Peter Checkland's Soft System Methodology (SSM), software designers are not allowed to assume that properties of the description always have a physical counterpart (Checkland 1981). Instead, relationships between features of the software design and reality are always a matter of conjecture, and always open to challenge. We will say that design methods that fall into this strand of research are **holistic** design methods.

A crude classification of software design methods by philosophical inclination produces the four categories, shown in Table 1; examples of the design methods associated with each category are given in Table 2.

| Research Strand | Epistemological Position | Ontological Position |
|---|---|---|
| Formal | Rationalist | Realist |
| Semi-Formal | Rationalist | Anti-Realist |
| Object-Oriented | Empiricist | Realist |
| Holistic | Empiricist | Anti-Realist |

**Table 1 – Philosophical Positions by Research Strand**

| Research Strand | Design Methods |
|---|---|
| Formal | Unity, Z, VDM |
| Semi-Formal | Jackson System Development |
| Object-Oriented | Booch Object Oriented Design |
| Holistic | Soft Systems, Multiview |

**Table 2 – Design Methods by Research Strand**

We recognise that finer points of distinction could be made and that ontological and epistemological arguments are rarely orthogonal, nonetheless we claim the distinctions in Table 1 are sufficiently fine to act as valid foundations for different ad hoc theories of software.

## 1.4 Theories of Software

By a theory of software, we mean a theory capable of characterising the descriptions used in software design. In program design, an underlying theory of programming (i.e. a theory of computation) characterises all possible valid programs. By tradition, the formal theory of programs is the only theory recognised as a valid foundation for software design.

A key problem for a theory of software design however, is that each of the philosophical positions described above has a *distinct* notion of equivalence. In King D. and Kimble C. (2004), we consider the precise differences in the notion of equivalence between the four positions, but in this paper, we will take equivalence to mean an isomorphic relationship between the things at either side of the relationship. Thus, for example, equivalence (denoted by '≡' in figure 2) between description and representation means things can be mapped between the description and representation, without changing the meaning of the relationship. This notion of equivalence characterises a rationalist position where no distinction exists between descriptions and representations. In contrast, empiricist positions make a distinction between observations and the recording of observations: description and representation are not isomorphic, and so not equivalent.

Looking at figure 2 overleaf, we can see a summary of the different equivalence relationships between descriptions, representations and reality for each of the four strands. Each of the four philosophical positions is fundamentally different and none can easily be reconciled with the others. Consequently, rather than attempting to define a single, unified theory of software design, we advocate a move towards an ad hoc theory of design where a particular theory is devoted to a particular purpose to which it is most suited.

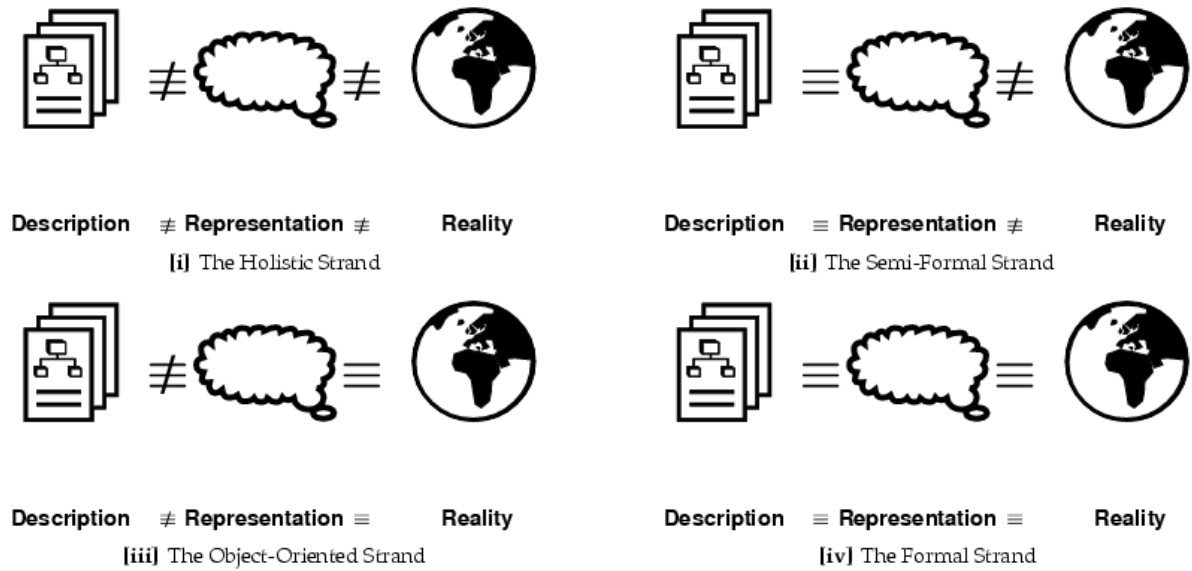

Figure 2 – Notions of equivalence in different ad hoc theories of software

## 1.5 The move to ad hoc theories

The influence of formal strand of software design is so powerful and all pervasive that software designers are reluctant to *explicitly* create a software design method that does not have a notion of equivalence similar to that found in formal theories. However, this frequently means software design methods attempt to retrofit other, inappropriate, notions of equivalence onto formal theories, which inevitably leads to problems.

We argue that this effort is misplaced for the following reasons.

1. **The concepts of 'software' and 'programs' are distinct.** Although the term's 'software' and 'program' are often used interchangeably, most researchers draw some distinction between 'software design' and 'program design'. In the early years, the problems faced by designers focused clearly on the machine. However, as problems became larger designers quickly realised that the computer-oriented focus did not always help. Gradually these descriptions of the 'larger-purposes' of the program became known as software. Software design is now recognised as a broader activity than program design. Thus, there is good reason to expect that a theory of software (characterising all software design) will differ from a theory of programs (characterising all program design).

2. **Software designers already use ad hoc theories of software**. The language software designers choose to use to describe software matters because it must be able to describe the, often complicated, situations found in the real world. Languages that force the designer to spell out the relationship in order to enforce formal equivalence may simply end up introducing unnecessary detail into the description. For example, object-oriented design methods recognise that not all relationships between descriptions of entities and objects in the real world are static. Occasionally designers might discover new properties of objects in reality, forcing an update in the descriptions of entities. Object-oriented design methods therefore need (and use) a notion of "*equivalent until further notice*": a notion that does not exist in formal theory. Thus, although software design is often inspired by program design, the particular needs of software design force the designer to search for ad hoc solutions.

3. **No single position explains all behaviour of software.** Lehman (2001) examined the behaviour of several large software projects. He divides software projects between two categories of software. He calls software completely characterised by a formal specification of the software behaviour Specification-Type Software (S-Type Software). All other categories of software are placed in the category of Embedded-Type Software (E-Type Software), be-cause the software is embedded in an inherently non-formal world. For Lehman, all software should be S-Type, because formal theories form the central foundations for all program designs. Lehman hypothesises that as S-Type blocks of software as built together, the complexity of the world works to degrade the properties of the individual blocks and the software ceases to behave according to the demands of formal theory. Thus, as the behaviour of software does not conform to the behaviour expected of descriptions using only formal notions of equivalence, this diversity of behaviour argues for a series of ad hoc theories of software, rather than a single unified theory of software.

We have outlined our arguments against a unified theory of software design based on a single philosophical foundation and for the use of ad hoc theories, each taking a particular philosophical stand without making any claims to universality. Further, we argue that although, in principle, any ad hoc theory could be used in any software design process, in practice the assumptions of some ad hoc theories will prove to have a better match with the reality of a particular software design process than others. It is this argument that we examine using empirical data in the next section of this paper.

## 2. Empirical Testing

Although the existence of ad-hoc theories of software may be accepted conceptually, a major practical challenge is identifying the conditions where the assumptions of a particular ad hoc theory and the reality of a particular software design process are likely to match. Having briefly described the theoretical background to this work, the remainder of the paper will turn to the problem of evaluating this theoretical framework against the reality of software design.

## 2.1 Uncovering empirical foundations

One way of uncovering the conditions where the assumptions of a particular ad hoc theory might apply is to look for empirical support. We can do this by taking the philosophical positions in Table 1 and interpreting them at face value.

For example, take the relationship between the description and reality. Both formal and semi-formal methods of software design take a rationalist epistemological position and argue that reality is governed by an immutable set of rules. Thus, reality forms a stable base for the software design. Formal methods of software design go further, adopting a realist ontological position and arguing that the description is also governed by an immutable set of rules linked to reality. Semi-formal methods however do not go this far and argue that while reality is governed by an immutable set of rules, designers are not always perfect at describing those rules.

So, although both formal and semi-formal design methods see reality as being fixed, differences exist in the assumptions made about the way in which reality can be described. While semi-formal design methods assume the description will inevitably change as more is learnt about a particular problem, formal design methods assume that any 'errors' in the description can be discovered and eliminated. Formal design methods assume that all 'errors' in a description are trivial, because errors only occur when rules are misapplied. By contrast, semi-formal design methods allow not only for the misapplication of rules, but also for the rules themselves to be flawed. In other words, there is an assumption that descriptions in formal design methods are free of major errors while in semi-formal design methods descriptions are expected to contain non-trivial errors that will need to be corrected.

By examining data from real software projects, we should be able to determine when these assumptions apply. This in turn should allow us to make an empirical assessment of the applicability of a particular ad hoc theory of software to a particular set of circumstances. The next section of this paper presents the results of one attempt to do just this using data from three large software development projects.

## 2.2 An Example of Empirical Testing
### 2.2.1  Background

This example is based on three separate projects from Rolls Royce. Each project is concerned with the development of software for a control system for the same class of complex component. Each component type has a number of distinct sub-types and the development of the software for each particular sub-type is tracked.

The development of software for control systems is a well established and refined science. The core principals and control laws are well known and the designers of control systems have access to a wealth of well documented engineering expertise. Although the basic pattern for the control system is well known, the demands of specific customers vary; hence the need for specific sub-types.

Within each sub-project, development is divided into a number of stages. These stages are not always linearly ordered; frequently pressures of time mean there is a considerable overlap between the stages. Notwithstanding this, each stage is effectively a full software development cycle. New requirements can be fed into a stage, assumptions changed and test results fed back. Once the changes are made, the results are handed on to another stage, either during, or at the end of the stage, depending on the circumstances.

All three projects in this example were developed using design methods following what we have characterised as the semi-formal approach. As previously noted, Rolls-

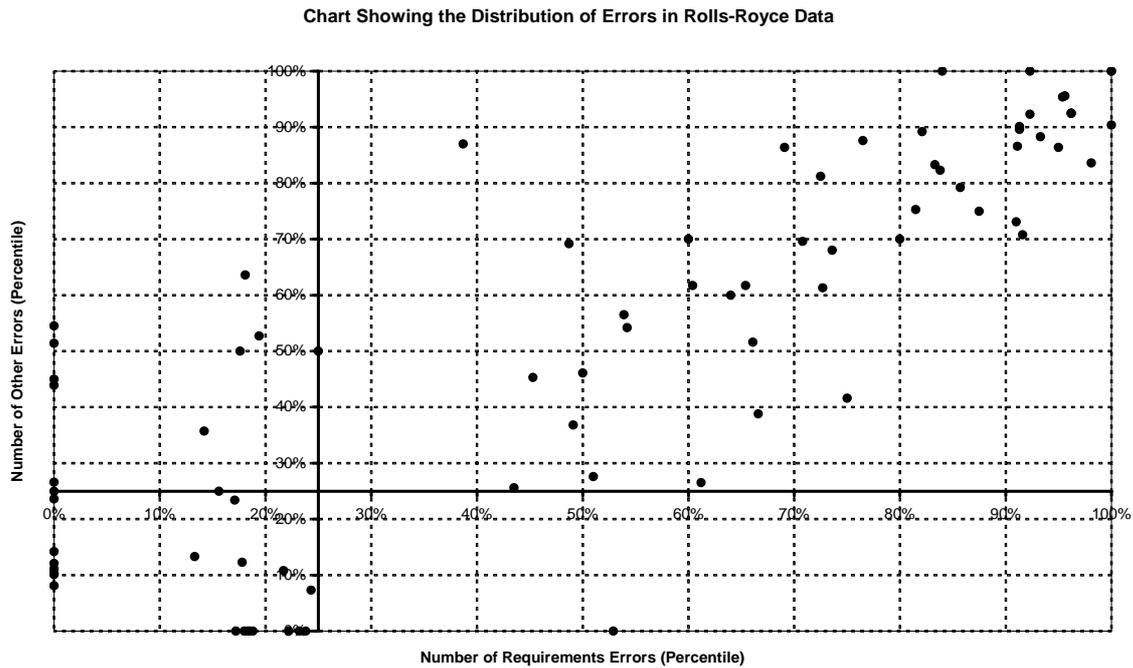

**Figure 3 – Distribution of Errors (by Percentile) in Rolls-Royce Data**

Royce has well developed engineering expertise in both the theoretical and practical design of control systems; consequently, we would expect the assumptions of the semi-formal design methods to be a good match for this particular example of software development.

### 2.2.2 Checking Assumptions

Following from our previous discussion, if a semi-formal approach to design was adopted, we would expect the descriptions to be stable, although the context of the description (the designer's representation of reality) would change.

To turn these rather general assumptions into a set of criteria suitable for empirical testing we focused on two metrics for which data was readily available: errors in software requirements ($E_R$) and other errors in the software description ($E_O$). We then used the ratio of requirements errors to other errors as a simple metric for uncovering whether the assumptions of the semi-formal design method held.

For a semi-formal design method, if the descriptions are stable, then any errors in the descriptions should be rare. However, when the context of the description is unstable, then formulating stable requirements is difficult and we expect to see a number of requirements errors. Thus in the unstable conditions of real software design projects, we would expect "other errors" to be rare (i.e. $E_O \approx 0$) with most errors occurring errors in the specification of requirements (i.e. $E_R > 0$).

The positions of the other three design methods can be deduced by following similar reasoning and are summarised in Table 3 below.

| Research Strand | $E_R$ | $E_O$ |
|---|---|---|
| Formal | $\approx 0$ | $\approx 0$ |
| Semi-Formal | $\approx 0$ | $> 0$ |
| Object-Oriented | $> 0$ | $\approx 0$ |
| Holistic | $> 0$ | $> 0$ |

**Table 3 – Assumptions by Software Research Strand**

### 2.2.3 Results

Figure 3 shows the results from all three projects, in the form of a scatter plot. To construct this diagram the data for the number of requirements and other errors was divided into percentiles and ranked. For both the number of requirements and other errors, values in the first quartile were taken to be $\approx 0$. Values ranked at, or above the $25^{th}$ percentile were taken to be $> 0$. Thus, the X and Y axes in the graph shown in Figure 3 are drawn at the $25^{th}$ percentiles.

Looking at Figure 3, we can see that most values lie above the $25^{th}$ percentiles. We can therefore conclude that for most development stages both $E_R > 0$ and $E_O > 0$. This is a somewhat surprising result as this pattern fits the assumptions of the holistic design method better than the of the semi-formal design method that was supposed to be being used.

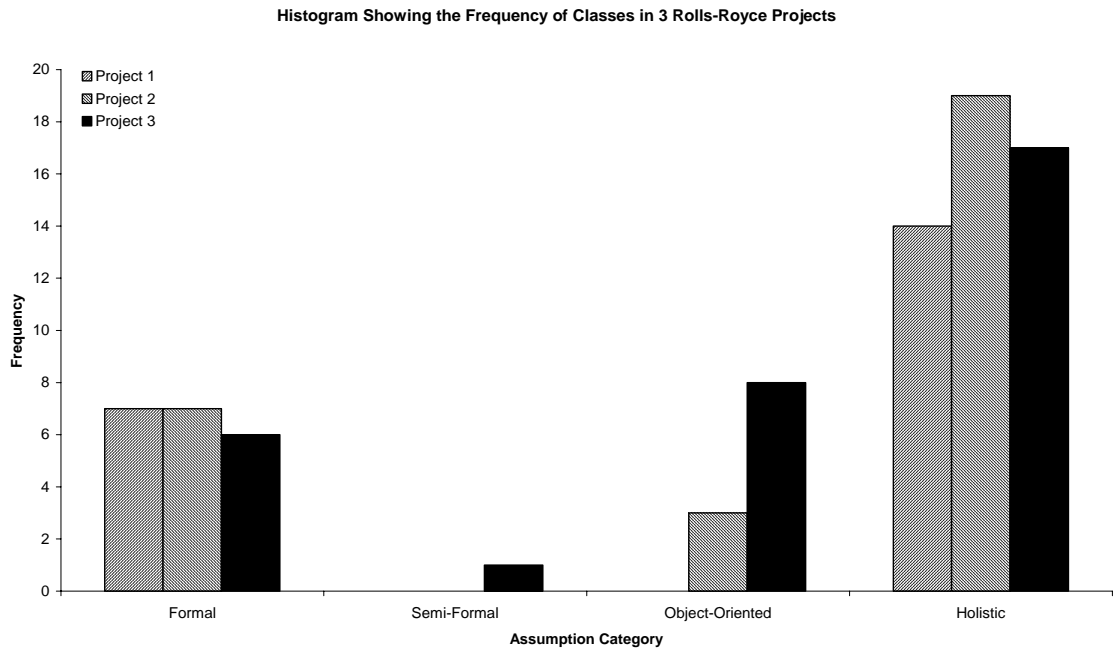

**Figure 4 – Frequency of Assumption Categories in the Example Rolls-Royce Data**

We can show the breakdown of the matches for the four sets of assumptions underpinning different design methods more clearly by simply classifying each development stage according to which box it fits in Figures 3. Applying this simple classification to each of the three projects gives us the results shown in Figure 4.

This shows more clearly which stages of the design process match with which set of assumptions. Thus, stages from the top right hand box (X and Y over the 25$^{th}$ percentile) can be tagged holistic. Similarly, stages in the bottom left hand box (X and Y under the 25$^{th}$ percentile) can be tagged as best suited to the assumptions of the formal design methods. We can now see clearly that the majority of stages actually followed the assumptions of the holistic design methods, with only one stage in one project following the assumptions of the semi-formal design methods that were supposed to be being used.

Interestingly, for both Project 1 and Project 2, most stages were either stable (which would follow the assumptions of the formal design methods), or unstable (which would follow the assumptions of the holistic design methods).

## 3. Conclusions

Focusing on the differences between the different philosophical positions underpinning different software design methods is useful: both from a practical and theoretical perspective. Most discussions of software design methods have aimed to establish a single, unified theory of software. These unified theories broadly follow the philosophical assumptions of formal programming theory. By making a distinction between software and programs, we draw two conclusions. Firstly, in reality, software designers adopt a broad range of philosophical positions most of which are not trivially reconcilable to those underpinning formal programming theory. Secondly, we can see no single philosophical position that exactly characterises the empirical behaviour of software development. This pattern of changing assumptions is repeated at a finer grained analysis of the software design process. For example, by examining each of the stages in the above examples we have shown that even characterising the development of a single stage in terms of one set of philosophical assumptions is difficult.

Some philosophical assumptions however appear to be more widely applicable than others are. For instance, patterns of behaviour we characterised as following the holistic position are the most common. Based on this work, it seems unlikely that, for the majority of software design processes, a single set of assumptions will ever be applicable.

One question that is not answered by this work is how far software designers can force the application of a particular set of assumptions. For instance, how far can software designers stabilise the requirements to favour the assumptions they need to make to use an object-oriented or formal design method? Successful software design projects may simply have succeeded in forcing the application of particular methods of software design!

Whether or not a design method can direct the thinking of a designer, the designers must be continuously aware of the assumptions they are making. We have seen that at some point the assumptions made by the designer and those made by the design method are likely to diverge. Whether this divergence is significant will depend on individual circumstances. Nonetheless, by embracing ad hoc theories (with narrow, distinct philosophical assumptions) designers are more likely to become aware of potential problems. By preserving the ability of designers to criticise their own assumptions, designers should gain a broader understanding of the strengths and weaknesses of particular methods. Backing this understanding with empirical studies should help to move the discipline of Information Systems onto a much firmer philosophical foundation. Rather than weakening the understanding of software, we believe that ad hoc theories offer a way towards a stronger and more grounded future for software design.

# References

Audi R. (1995), "The Cambridge Dictionary of Philosophy", Cambridge University Press, ISBN 0-521-40224-7.

Checkland P. (1981), "Systems Thinking, Systems Practice", John Wiley & Sons Ltd., ISBN 0-471-27911-0

Duro P. Greenhalgh M. Griffiths T. McLeish K. and O'Leary B. (1993), "Realism", in McLeish K. (1993).

Graham T. (1996), "Viewpoints Supporting the Development of Interactive Software", Joint Proceedings of the Second International Software Architecture Workshop (ISAW-2) and International Workshop on Multiple Perspectives in Software Development (Viewpoints '96) on SIGSOFT '96 Workshops", San Francisco, California, United States, ISBN 0-89-791867-3, pp. 263—267.

Jack A. (1993), "Rationalism", in McLeish K. (1993).

Kant I. (1996), "Critique of Pure Reason", Unified edition, Hackett Publishing Company Inc, ISBN 0-87220-258-5

King D. and Kimble C. (2004), "Notions of Equivalence in Software Design", 9$^{th}$ AIM Conference: Critical Perspectives on Information Systems, INT Evry, France, 2004

Lehman M. and Ramil J. (2001), "Rules and Tools for Software Evolution Planning and Management", Annals of Software Engineering, November, 11 (1), pp. 15–44.

Meadow C. and Yuan W. (1997), "Measuring the Impact of Information: Defining the Concepts", Information Processing & Management, November, 33 (6), pp. 697–714.

Simpson J. and Weiner E. (1989), "The Oxford English Dictionary", 2$^{nd}$ edition, Oxford University Press, Oxford, ISBN 0-19-861186-2.

Sommerville I. and Sawyer P. (1997), "Viewpoints: Principles, Problems and a Practical Approach to Requirements Engineering", Annals of Software Engineering, 3, ISSN 1022-7091, pp. 101--130.

Zúñiga G. (2001), "Ontology: Its Transformation from Philosophy to Information Systems", Proceedings of the International Conference on Formal Ontology in Information Systems, Ogunquit, Maine, USA, ISBN 1-58-113377-4, pp. 187—197.